\documentclass[prd,twocolumn,showpacs,preprintnumbers,amsmath,amssymb,superscriptaddress,floatfix,nofootinbib]{revtex4}

\usepackage{graphicx}
\usepackage{epstopdf}
\usepackage{bm}
\usepackage{amsmath}
\usepackage{amsfonts}
\usepackage{amssymb}
\usepackage{color}
\usepackage{multirow}
\usepackage{footnote}

\usepackage[colorlinks, citecolor=blue,anchorcolor=red,menucolor=red,linkcolor=red,filecolor=red,runcolor=red,urlcolor=blue,frenchlinks=red]{hyperref}
\begin{document}
\title{Coupled channel effects for the bottom-strange mesons}
\author{Wei Hao}
\email{haowei@nankai.edu.cn}
\affiliation{School of Physics, Nankai University, Tianjin 300071, China}

\author{M. Atif Sultan}%
\email{atifsultan.chep@pu.edu.pk}
\affiliation{School of Physics, Nankai University, Tianjin 300071, China}
\affiliation{Centre  For  High  Energy  Physics,  University  of  the  Punjab,  Lahore  (54590),  Pakistan}

\author{En Wang}~\email{wangen@zzu.edu.cn}
\affiliation{School of Physics, Zhengzhou University, Zhengzhou 450001, China}

\begin{abstract}
We have calculated the mass spectrum of $B_s$ mesons within a nonrelativistic potential model considering coupled channel effects, and the corresponding strong decay widths within the $^3P_0$ model using the numerically calculated wave functions. By comparing with the available experimental data, we find that the states $B_s$, $B_s^*$, $B_{s1}(5830)$, and $B_{s2}^*(5840)$ could be interpreted as the $B_s(1^1S_0)$, $B_s(1^3S_1)$, $B_s(1P^\prime)$, and $B_s(1^3P_2)$, respectively. Although the quantum numbers of the newly observed $B_s(6064)$ and $B_s(6158)$ states have not been determined, our results support the assignments of $B_s(1^3D_3)$ and $B_s(1^3D_1)$ for them. Our predictions are helpful in searching for the bottom-strange meson in future experiments.
\end{abstract}

\maketitle

\section{Introduction}
In past years, significant progress has been made in studying the bottom-strange mesons. Up to now, eight bottom-strange mesons have been observed in experiments, as shown in Table~\ref{tab:exp}. The ground state $B_s$ meson was first observed in 1990 using the Columbia University–Stony Brook (CUSB-II) detector~\cite{Lee-Franzini:1990dev}, and was confirmed by many collaborations, such as CDF, OPAL, Belle,  DELPHI, and LHCb~\cite{ParticleDataGroup:2024cfk}.  The $B_{s1}(5830)$ and $B_{s2}^*(5840)$ were first observed by the CDF Collaboration in the $K^-B^+$ two-body decay system~\cite{CDF:2007avt} and confirmed by LHCb~\cite{LHCb:2012iuq}, CMS~\cite{CMS:2018wcx}, and D0~\cite{D0:2007die}. However, the $B_{sJ}^*(5850)$ was only observed by the OPAL Collaboration at LEP in the $B^{(*)+}K^-$ invariant mass distribution in 1995~\cite{OPAL:1994hqv}, and more experimental information is needed to confirm the existence of this state. Until 2020, the $B_{sJ}(6064)$, $B_{sJ}(6114)$, $B_{sJ}(6109)$ and $B_{sJ}(6158)$ were founded in the $B^\pm K^\mp$ mass spectrum of the proton-proton collisions by LHCb~\cite{LHCb:2020pet}, which further enriches experimental information on the bottom-strange mesons. It should be pointed out that these four states may be just two states. Because the $B_{sJ}(6064)$ and $B_{sJ}(6114)$ are obtained by analyzing the decay directly into finial state $B^\pm K^\mp$. However, if the decay occurs through $B^{*\pm} K^\mp$ with a missing photon from the decay $B^{*\pm}\to B^\pm\gamma$ then the masses of the two states will shift approximately 45~MeV, which leads to the $B_{sJ}(6109)$ and $B_{sJ}(6158)$ states~\cite{LHCb:2020pet}.

\begin{table*}[!htpb]
\begin{center}
\caption{ \label{tab:exp} Experimental information of the $B_s$ mesons~\citep{ParticleDataGroup:2024cfk}.}
\footnotesize
\setlength{\tabcolsep}{1mm}{
\begin{tabular}{cccccc}
\hline\hline
  state               &mass  (MeV)              &width  (MeV)      & $I(J^{P})$                  &mode \\\hline
  $B_s$               &$5366.93\pm0.10$         &$-$               & $0(0^-)$                    &  \\ 
  $B_s^*$             &$5415.4\pm1.4$           &$-$               & $0(1^-)$                    &  \\
  $B_{s1}(5830)$      &$5828.73\pm 0.20$        &$0.5\pm0.4$       & $0(1^+)$                    & $B^{*+}K^-$  \\
  $B_{s2}^*(5840)$    &$5839.88\pm0.12$         &$1.49\pm0.27$     & $0(2^+)$                    & $B^+K^-, B^{*+}K^-$, $B^0K^0_S$, $B^{*0}K^0_S$  \\
  $B_{sJ}(6064)$      &$6063.5\pm1.2\pm0.8$     &$26\pm4\pm4$      & $0(?^?)$\cite{LHCb:2020pet} & $B^+K^-$    \\
  $B_{sJ}(6114)$      &$6114\pm3\pm5$           &$66\pm18\pm21$    & $0(?^?)$\cite{LHCb:2020pet} & $B^+K^-$   \\
  $B_{sJ}(6109)$      &$6108.8\pm1.1\pm0.7$     &$22\pm5\pm4$      & $0(?^?)$\cite{LHCb:2020pet} & $B^+K^-$   \\
  $B_{sJ}(6158)$      &$6158\pm4\pm5$           &$72\pm18\pm25$    & $0(?^?)$\cite{LHCb:2020pet} & $B^+K^-$    \\
  \hline\hline
\end{tabular}}
\end{center}
\end{table*}

Although a few bottom-strange states are observed experimentally, many theoretical approaches have been used to study bottom-strange mesons, such as the constituent quark model~\cite{Godfrey:1985xj,Zeng:1994vj,DiPierro:2001dwf,Ebert:2009ua,Lahde:1999ih,Lu:2016bbk,Hao:2022ibj,Feng:2022hwq,Li:2021qgz}, chiral quark model~\cite{Zhong:2008kd}, lattice QCD~\cite{Lang:2015hza}, heavy meson effective theory~\cite{Xu:2014mka,Wang:2014cta}. However, such traditional approaches cannot explain well the properties of some excited mesons, such as $X(3872)$, $D^*_{s0}(2317)$, and $D_{s1}(2460)$, which indicates that traditional approaches need some necessary modifications. For example, conventional quark models may miss the generation of the light quark-antiquark pair, which could enlarge the Fock space of the initial state~\cite{Lu:2016mbb}, and such effects will change the Hamiltonian of the conventional quark potential models and lead to mass shifts.
Thus, the unquenched quark models, accounting for such effects, have been developed, and there are two kinds of unquenched quark models, one is the screened potential models~\cite{Song:2015fha,Song:2015nia,Li:2009zu,Hao:2022ibj,Feng:2022esz,Hao:2024gud}, and the other is the coupled channel model~\cite{Yang:2023tvc,Ortega:2016mms,Hao:2022vwt,Ferretti:2013faa,Ferretti:2012zz,Ferretti:2013vua,Ferretti:2015rsa,Lu:2016mbb,Kalashnikova:2005ui,Hao:2024nqb,Hao:2024ptu,Pan:2016bac}, both of which have achieved great success in the interpretation of the hadron spectra. Within the unquenched quark models, the properties of the $X(3872)$, $D^*_{s0}(2317)$, $D_{s1}(2460)$ can be well explained~\cite{Eichten:2004uh,vanBeveren:2003kd,Dai:2006uz,Liu:2009uz,Ortega:2016mms,Hao:2022vwt,Cao:2024nxm}. Thus, to describe the newly observed bottom-strange mesons $B_{sJ}(6064)$, $B_{sJ}(6114)$, $B_{sJ}(6109)$ and $B_{sJ}(6158)$ and predict the accurate masses and decay properties ofa the excited bottom-strange mesons, in this work we will systematically investigate the mass spectrum and decay properties of the bottom-strange mesons within the coupled channel model and discuss the possible assignments of these states.

The paper is organized as follows. In Section~\ref{sec:formalism}, we will introduce the coupled channel framework. The numerical results and the corresponding discussions are presented in Section~\ref{sec:results}. Finally, a short summary will be given in Section~\ref{sec:summary}.

\section{Theoretical formalism}\label{sec:formalism}
\subsection{Coupled channel framework}
In the coupled channel model, the full Hamiltonian is composed of quenched part and self energy part, and can be expressed as  
\begin{equation} \label{eqn:full}
	\mathcal{H} = \mathcal{H}_{b\bar{s}} + \mathcal{H}_{BC} + \mathcal{H}_I,
\end{equation}
where the quenched part labeled as $\mathcal{H}_{b\bar{s}}$ accounts for the interaction between constituent quarks $b$ and $\bar{s}$, and the self energy part includes the energy term $\mathcal{H}_{BC}$ and the mixing term $\mathcal{H}_I$. The term $\mathcal{H}_{BC}$ can be written as $\mathcal{H}_{BC} = E_{BC} =\sqrt{m_B^2 +p^2} + \sqrt{m_C^2 +p^2}$, where $B$ and $C$ stand for two mesons of the coupled channel. 
With the Hamiltonian, the physical masses of the bottom-strange mesons can be written as
\begin{equation}
\label{m}
M = M_{b\bar{s}} + \Delta M, 
\end{equation}
with
\begin{equation}
\Delta M = \sum_{BC,\ell,S} \int_0^{\infty} p^2 dp \frac{\left|\left\langle BC;p ,\ell S\right| \mathcal{H}_I \left| \psi_0 \right\rangle \right|^2}{M - E_{BC}},
\end{equation}
where $M_{b\bar{s}}$ is the bare mass, and $\Delta M$ is the mass shift from the coupled channel effects. Here $p$ is the momentum of the one meson in the $BC$ system. We will give the expression of the $\mathcal{H}_I$ in the following.  $\ell$ is the relative angular momentum of $B$ and $C$, and  $S$ is the sum of the $B$ and $C$ spins.

Besides the mass shift, the probabilities of the various components can be obtained by analyzing the wave functions
\begin{align} \label{eqn:psi}
    |\psi\rangle = c_0 |\psi_0\rangle + \sum_{BC, \ell, S} \int d^3p\, c_{BC}(p) |BC;p, \ell S\rangle,
\end{align}
where the first term relates to $b\bar{s}$ component, and the second term relates to self energy component. By considering the normalization condition, $|c_0|^2+\int d^3p|c_{BC}|^2=1$, the probabilities of the $b\bar{s}$ component in the state below the $BC$ threshold can be written as 
\begin{eqnarray}
\label{eqn:pqqbar}
	\mathcal{P}_{b\bar{s}} &\equiv& |c_0|^2 \nonumber \\&=& \left(1+\sum_{BC,\ell,S} \int_0^{\infty} p^2 dp \frac{\left|\left\langle BC;p, \ell S \right| \mathcal{H}_I \left| \psi_0 \right\rangle \right|^2}{(M - E_{BC})^2}\right)^{-1},
\end{eqnarray}
and
\begin{eqnarray}
 \mathcal{P}_{BC}&= &  \left(1+\sum_{BC,\ell,S} \int_0^{\infty} p^2 dp \frac{\left|\left\langle BC;p ,\ell S \right|\mathcal{H}_I \left| \psi_0 \right\rangle \right|^2}{(M - E_{BC})^2}\right)^{-1}\nonumber \\
 &&\times \sum_{\ell,S} \int_0^{\infty} p^2 dp \frac{\left|\left\langle BC;p, \ell S \right| \mathcal{H}_I\left| \psi_0 \right\rangle \right|^2}{(M - E_{BC})^2}.
\end{eqnarray}

Furthermore, for the initial state above the $BC$ threshold, the strong decay width can be obtained by the mixing term $\mathcal{H}_I$ as
\begin{equation}
\label{eqn:decay}
    \Gamma_{BC} = 2 \pi p_0 \frac{E_B(p_0) E_C(p_0)}{M} \sum_{\ell,S} \left| \left\langle BC;p_0, \ell S\right| \mathcal{H}_I \left| \psi_0 \right\rangle \right|^2,
\end{equation}
where $p_0$ is the momentum of the final meson in the rest frame of the initial state, and $E_B$ and $E_C$ are the energy of the $B$ and $C$ mesons, respectively,

\subsection{Bare mass}
The bare mass $M_0$ corresponds to the quenched part of the coupled channel model, and could be obtained within the potential models,  such as the Godfrey-Isgur (GI) relativized quark model~\cite{Godfrey:1985xj}, constituent quark model~\cite{Vijande:2004he}, nonrelativistic quark model~\cite{Lakhina:2006fy}. Since the botton-strange mesons could be regarded as the nonrelativistic system due to the large mass of the bottom quark, in this work we use the nonrelativistic quark model to calculate the bare mass, which has been successfully used to describe various mesons ~\cite{Lu:2016bbk,Li:2010vx,Hao:2019fjg,Feng:2022esz}. The Hamiltonian of the nonrelativistic quark model is denoted by $\mathcal{H}_{b\bar{s}}$, 
\begin{align}
    \mathcal{H}_{b\bar{s}} =& \mathcal{H}_0+\mathcal{H}_{sd},\\
    \mathcal{H}_0 =& m_{b} + m_{\bar{s}} + \frac{\boldsymbol{P}^2}{2M_r}-\frac{4}{3}\frac{{\alpha}_s}{r}+br+C_{b\bar{s}}\nonumber\\  & + \frac{32{\alpha}_s{\sigma}^3 e^{-{\sigma}^2r^2}}{9\sqrt{\pi}m_bm_{\bar{s}}} {\boldsymbol{S}}_{b} \cdot {\boldsymbol{S}}_{\bar{s}},
\end{align}
where $m_{b}, m_{\bar{s}}$ are the masses of the constituent quarks, $\alpha_s$, $b$, and $\sigma$ are the free parameters which reflect the interaction strengths. $C_{b\bar{s}}$ is a constant that controls the overall movement of meson masses. $M_r=m_b m_{\bar{s}}/(m_b+m_{\bar{s}})$ is the reduced mass of the bottom and strange quarks. $\boldsymbol{S}_b$ and $\boldsymbol{S}_{\bar{s}}$ are the spin of the bottom and strange quarks, respectively. 

The spin-dependent term $\mathcal{H}_{sd}$ includes spin-orbit term and spin-spin term,
    \begin{eqnarray}
      \mathcal{H}_{sd} &=& \left(\frac{\boldsymbol{S}_{b}}{2m_b^2}+\frac{{\boldsymbol{S}}_{\bar{s}}}{2m_{\bar{s}}^2}\right) \cdot \boldsymbol{L}\left(\frac{1}{r}\frac{dV_c}{dr}+\frac{2}{r}\frac{dV_1}{dr}\right)\nonumber\\
      &&+\frac{{\boldsymbol{S}}_+ \cdot \boldsymbol{L}}{m_bm_{\bar{s}}}\left(\frac{1}{r} \frac{dV_2}{r}\right) \nonumber\\
      && +\frac{3{\boldsymbol{S}}_{b} \cdot \hat{\boldsymbol{r}}{\boldsymbol{S}}_{\bar{s}} \cdot \hat{\boldsymbol{r}}-{\boldsymbol{S}}_{b} \cdot {\boldsymbol{S}}_{\bar{s}}}{3m_bm_{\bar{s}}}V_3\nonumber\\
      && +\left[\left(\frac{{\boldsymbol{S}}_{b}}{m_b^2}-\frac{{\boldsymbol{S}}_{\bar{s}}}{m_{\bar{s}}^2}\right)+\frac{{\boldsymbol{S}}_-}{m_b m_{\bar{s}}}\right] \cdot \boldsymbol{L} V_4,
\end{eqnarray}
where $\boldsymbol{L}$ is the orbital angular momentum between quark $b$ and antiquark $\bar{s}$, and $\boldsymbol{S}_{\pm}={\boldsymbol{S}}_b\pm{\boldsymbol{S}}_{\bar{s}}$.  The $V_c$ and $V_1\sim V_4$ can be written as
\begin{eqnarray}
  V_c &=& -\frac{4}{3}\frac{{\alpha}_s}{r}+br,\nonumber \\
  V_1 &=& -br-\frac{2}{9\pi}\frac{{\alpha}_s^2}{r}[9{\rm ln}(\sqrt{m_bm_{\bar{s}}}r)+9{\gamma}_E-4],\nonumber\\
  V_2 &=& -\frac{4}{3}\frac{{\alpha}_s}{r}-\frac{1}{9\pi}\frac{{\alpha}_s^2}{r}[-18{\rm ln}(\sqrt{m_bm_{\bar{s}}}r)+54{\rm ln}(\mu r)\nonumber\\
  &&+36{\gamma}_E+29],\nonumber\\
  V_3 &=& -\frac{4{\alpha}_s}{r^3}-\frac{1}{3\pi}\frac{{\alpha}_s^2}{r^3}[-36{\rm ln}(\sqrt{m_bm_{\bar{s}}}r)+54{\rm ln}(\mu r)\nonumber\\
  &&+18{\gamma}_E+31],\nonumber\\
  V_4 &=& \frac{1}{\pi}\frac{{\alpha}_s^2}{r^3}{\rm ln}\left(\frac{m_{\bar{s}}}{m_b}\right),
\end{eqnarray}
where $\mu=1$~GeV is renormalization scale, $\gamma_E=0.5772$ is the Euler constant.

Since the $b\bar{s}$ mesons are not charge conjugation eigenstates,  which means the mixing will occur only when states have $J=L$ and $S=0,1$. For example, the two $J^P=1^+$ states $1^1P_1$ and $1^3P_1$ can mix into $1P$ and $1P^\prime$ with mixing angle $\theta$. The mixing angle can be obtained by analyzing the spin-orbit term $\mathcal{H}_{sd}$, which could be divided  into symmetric part $\mathcal{H}_{sym}$ and antisymmetric part $\mathcal{H}_{anti}$~\cite{Lu:2016bbk},
\begin{eqnarray}
\mathcal{H}_{sym} &=& \frac{{\boldsymbol{S}}_+ \cdot {\boldsymbol{L}}}{2}\left[\left(\frac{1}{2m_b^2}+\frac{1}{2m_{\bar{s}}^2}\right) \left(\frac{1}{r}\frac{dV_c}{dr}+\frac{2}{r}\frac{dV_1}{dr}\right)\right. \nonumber \\
&& \left.+\frac{2}{m_bm_{\bar{s}}}\left(\frac{1}{r} \frac{dV_2}{r}\right)+\left(\frac{1}{m_b^2}-\frac{1}{m_{\bar{s}}^2}\right)V_4\right],
\end{eqnarray}
\begin{eqnarray}
\mathcal{H}_{anti} &=& \frac{{\boldsymbol{S}}_- \cdot {\boldsymbol{L}}}{2}\left[\left(\frac{1}{2m_b^2}-\frac{1}{2m_{\bar{s}}^2}\right) \left(\frac{1}{r}\frac{dV_c}{dr}+\frac{2}{r}\frac{dV_1}{dr}\right)\right. \nonumber \\
&& \left.+\left(\frac{1}{m_b^2}+\frac{1}{m_{\bar{s}}^2}+\frac{2}{m_bm_{\bar{s}}}\right)V_4\right].
\end{eqnarray}
The antisymmetric part $\mathcal{H}_{anti}$  can provide non-diagonal terms in the matrix equation and then the mixing angle can be extracted by diagonalization calculation.
In this case, the the $B_s(nL)$, and $B_s(nL^\prime)$ states can be obtained as~\cite{Godfrey:1985xj,Godfrey:1986wj,Lu:2016bbk},
\begin{equation}
\left(
\begin{array}{cr}
B_{s L}(nL)\\
B_{s L}(nL^\prime)
\end{array}
\right)
 =\left(
 \begin{array}{cr}
\cos \theta_{nL} & \sin \theta_{nL} \\
-\sin \theta_{nL} & \cos \theta_{nL}
\end{array}
\right)
\left(\begin{array}{cr}
B_s(n^1L_L)\\
B_s(n^3L_L)
\end{array}
\right).
\label{eqn:mix}
\end{equation}

\subsection{Mass shift}
The mass shift comes from the transition between the physical state and coupled channel, which could be described by the mixing term Hamiltonian $\mathcal{H}_I$. Usually, this mixing is achieved by the $^3P_0$ model, which is a useful tool to analyze the coupling between the initial state and coupled channel~\cite{Micu:1968mk, LeYaouanc:1972vsx, LeYaouanc:1973ldf,Li:2022ybj,Xue:2018jvi,Wang:2017pxm}. In the  $^3P_0$ model, the quarks of the initial state can connect with vacuum generated quark-antiquark pair to form finial states, and the transition operator $\mathcal{H}_I=T^\dagger$ is expressed as~\cite{Ferretti:2013faa,Ferretti:2012zz,Ferretti:2013vua},
\begin{equation}
	\label{eqn:Tdag}
	\begin{array}{rcl}
	T^{\dagger} &=& -3 \, \gamma_0^{eff} \, \int d \vec{p}_3 \, d \vec{p}_4 \, 
	\delta(\vec{p}_3 + \vec{p}_4) \, C_{34} \, F_{34} \,  
	{e}^{-r_q^2 (\vec{p}_3 - \vec{p}_4)^2/6 }\,  \\
	& & \left[ \chi_{34} \, \times \, {\cal Y}_{1}(\vec{p}_3 - \vec{p}_4) \right]^{(0)}_0 \, 
	b_3^{\dagger}(\vec{p}_3) \, d_4^{\dagger}(\vec{p}_4) ~,   
	\end{array}
\end{equation}
where the generated quark-antiquark pair has vacuum quantum numbers $J^{PC} = 0^{++}$.
It should be noted that the operator $T^{\dagger}$ includes a Gaussian factor that represents the actual size of the created quark pair, which has been adopted to analyze meson properties~\cite{Ferretti:2013faa,Ferretti:2012zz,Ferretti:2013vua,Silvestre-Brac:1991qqx,Geiger:1991ab,Geiger:1991qe,Geiger:1996re}. It is believed that, when one sums over a complete set of virtual decay channels, this factor is necessary~\cite{Geiger:1991ab}. 
Here we use $r_q = 0.3$~fm, the middle value of the range $0.25$ – $0.35$~fm determined in Refs.~\cite{Silvestre-Brac:1991qqx,Geiger:1991ab,Geiger:1991qe,Geiger:1996re}. In Eq.~(\ref{eqn:Tdag}), the color, flavor, and spin wave functions of the $b\bar{s}$ meson system is labeled as $C_{34}$, $F_{34}$, and $\chi_{34}$, respectively. $ b_3^{\dagger}(\vec{p}_3)$ and $d_4^{\dagger}(\vec{p}_4)$ are the creation operators for a quark and an antiquark with momenta $\vec{p}_3$ and $\vec{p}_4$, respectively.  
For $u\bar{u}/d\bar{d}$ pair creation, the pair-creation strength $\gamma_0^{eff}=\gamma_0$ and for $s\bar{s}$ pairs, the $\gamma_0^{eff}=\frac{m_u}{m_s}\gamma_0$. Usually, the parameter $\gamma_0$ is determined by analyzing the decay properties of the mesons, and in this work we take a typical value $\gamma_0=0.4$, which has been used in many works to calculate various meson strong decay widths~\cite{Ackleh:1996yt,Barnes:2002mu,Barnes:2005pb,Close:2005se,Li:2019tbn,Godfrey:2016nwn,Godfrey:2015dva}.

\section{Results and discussions}
\label{sec:results}
\begin{table}[h] 
\caption{Parameters used in this work~\cite{Lakhina:2006fy,Li:2010vx,Lu:2016bbk}.} 
\label{tab:para}
\begin{center}
\begin{tabular}{ccc} 
\hline 
\hline 
Parameter  &  value \\
\hline
$m_n$      & $0.45$ GeV \\
$m_s$      & $0.55$ GeV\\
$m_b$      & $4.5$ GeV\\   
$\alpha_s$ & $0.5$  \\  
$b$        & $0.14$ GeV$^2$\\  
$\sigma$   & $1.17$ GeV \\
$C_{bs}$   & $0.169$ GeV \\  
$\gamma_0$ & $0.4$  \\
\hline 
\hline
\end{tabular}
\end{center}
\end{table}

\begin{table*}[htpb]
\begin{center}
\caption{\label{tab:spectrum} The mass spectrum (in MeV) of the $b\bar{s}$ mesons.
Column 3 to 5 stand for bare mass from the potential model, the mass shift, and the spectrum with coupled channel effects.
Results from other models are listed in Columns $6\sim9$ as comparison.
The last Column is the experimental values taken from RPP \cite{ParticleDataGroup:2022pth}. The mixing angles of $1P$, $2P$, $1D$ are $-55.8^\circ$, $-54.5^\circ$ and $-50.3^\circ$, respectively.
}
\footnotesize
\begin{tabular}{cccccccccccc}
\hline\hline
  $n^{2S+1}L_J$  & state            &$M_0$  &$\Delta M$  &$M$  &MGI~\cite{Hao:2022ibj}   &NR\cite{Lu:2016bbk}    &GI\cite{Godfrey:2016nwn}     &DRV\cite{Ebert:2009ua}    &KA\cite{Gandhi:2022nnk} &VRA\cite{Patel:2022hhl}   & RPP~\cite{ParticleDataGroup:2024cfk}  \\\hline
  $1^1S_0$     & $B_s$              &$5480$   &$-114$     &$5367$   &5367      &5362   &5394   &5372   &       &5359  &$5366.93\pm0.10$    \\
  $1^3S_1$     & $B_s^{*}$          &$5531$   &$-122$     &$5409$   &5419      &5413   &5450   &5414   &       &5415  &$5415.4\pm1.4$    \\
  $2^1S_0$     & $-$                &$6095$   &$-146$     &$5949$   &5947      &5977   &5984   &5976   &6025   &5980  &$-$           \\
  $2^3S_1$     & $-$                &$6122$   &$-130$     &$5992$   &5972      &6003   &6012   &5992   &6033   &5993  &$-$          \\
  $1^3P_0$     & $-$                &$5876$   &$-145$     &$5732$   &5753      &5756   &5831   &5833   &5709   &5798  &$-$      \\
  $1P$         & $-$                &$5921$   &$-145$     &$5776$   &5797      &5801   &5857   &5831   &5768   &5818  &$-$       \\
  $1P^\prime$  & $B_{s1}(5830)$     &$5953$   &$-147$     &$5805$   &5825      &5836   &5861   &5865   &5875   &5846  &$5828.73\pm 0.20$     \\
  $1^3P_2$     & $B_{s2}^*(5840)$   &$5968$   &$-143$     &$5825$   &5840      &5851   &5876   &5842   &5890   &5838  &$5839.88\pm0.12$     \\
  $2^3P_0$     & $-$                &$6323$   &$-110$     &$6214$   &6158      &6203   &6279   &6318   &6387   &6292  &$-$ \\
  $2P$         & $-$                &$6362$   &$-87$      &$6274$   &6194      &6241   &6279   &6321   &6393   &6304  &$-$  \\
  $2P^\prime$  & $-$                &$6413$   &$-106$     &$6307$   &6236      &6297   &6296   &6345   &6470   &6320  &$-$   \\
  $2^3P_2$     & $-$                &$6427$   &$-74$      &$6352$   &6249      &6309   &6295   &6359   &6476   &6316  &$-$ \\
  $1^3D_1$     & $B_{sJ}(6158)$     &$6261$   &$-104$     &$6157$   &6117      &6142   &6182   &6209   &6247   &6144  &$6158\pm4\pm5$    \\
  $1D$         & $-$                &$6205$   &$-128$     &$6077$   &6053      &6087   &6169   &6189   &6256   &6139  &$-$    \\
  $1D^\prime$  & $-$                &$6277$   &$-123$     &$6154$   &6132      &6159   &6196   &6218   &6292   &6135  &$-$    \\
  $1^3D_3$     & $B_{sJ}(6064)$     &$6214$   &$-135$     &$6079$   &6061      &6096   &6179   &6191   &6297   &6139  &$6063.5\pm1.2\pm0.8$       \\
  \hline\hline
\end{tabular}
\end{center}
\end{table*}

\begin{table*}
\caption{\label{tab:shift} Mass shift $\Delta M$ (in MeV) of each coupled channel.} 
\begin{tabular}{cccccccccccccc} 
\hline 
\hline 
State       &                   &$BK$   &$B^*K$ &$BK^*$ &$B^*K^*$   &$B_s\eta$  &$B_s\eta^\prime$ &$B_s^*\eta$ &$B_s^*\eta^\prime$ &$B_s\phi$ &$B_s^*\phi$  &Total \\ \\
\hline 
$1^1S_0$    &$B_s$              &$0$     &$-23$    &$-26$  &$-51$   &$0$    &$0$      &$-5$     &$-2$      &$0$     &$-6$    &$-114$        \\
$1^3S_1$    &$B_s^{*}$          &$-9$    &$-16$    &$-19$  &$-64$   &$-2$   &$-1$     &$-4$     &$-1$      &$-2$    &$-4$    &$-122$       \\
$2^1S_0$    &$-$                &$0$     &$-43$    &$-31$  &$-58$   &$0$    &$0$      &$-8$     &$-2$      &$0$     &$-5$    &$-146$    \\
$2^3S_1$    &$-$                &$-4$    &$-17$    &$-22$  &$-71$   &$-3$   &$-1$     &$-7$     &$-1$      &$-2$    &$-3$    &$-130$      \\    
$1^3P_0$    &$-$                &$-42$   &$0$      &$0$    &$-93$   &$-5$   &$-1$     &$0$      &$0$       &$-4$    &$0$     &$-145$     \\ 
$1P$        &$-$                &$0$     &$-42$    &$-30$  &$-59$   &$0$    &$0$      &$-6$     &$-2$      &$0$     &$-6$    &$-145$       \\     
$1P^\prime$ &$B_{s1}(5830)$     &$0$     &$-42$    &$-24$  &$-70$   &$0$    &$0$      &$-6$     &$-2$      &$0$     &$-5$    &$-147$    \\ 
$1^3P_2$    &$B_{s2}^*(5840)$   &$-17$   &$-23$    &$-23$  &$-63$   &$-3$   &$-1$     &$-4$     &$-2$      &$-3$    &$-4$    &$-143$      \\
$2^3P_0$    &$-$                &$-2$    &$0$      &$0$    &$-102$  &$-3$   &$-1$     &$0$      &$0$       &$-2$    &$0$     &$-110$     \\ 
$2P$        &$-$                &$0$     &$0.02$   &$-26$  &$-55$   &$0$    &$0$      &$-1$     &$-1$      &$0$     &$-4$    &$-87$     \\     
$2P^\prime$ &$-$                &$0$     &$-0.2$   &$-10$  &$-90$   &$0$    &$0$      &$-2$     &$-1$      &$0$     &$-3$    &$-106$    \\  
$2^3P_2$    &$-$                &$-4$    &$-3$    &$-13$   &$-47$   &$-0.2$ &$-1$     &$-0.2$   &$-1$      &$-2$    &$-3$    &$-74$     \\ 
$1^3D_1$    &$B_{sJ}(6158)$     &$12$    &$5$      &$-10$    &$-106$    &$-0.04$   &$-1$     &$-1$    &$-0.4$    &$-2$    &$-1$   &$-104$     \\
$1D$        &$-$                &$0$     &$-23$    &$-32$    &$-60$    &$0$    &$0$     &$-6$    &$-2$    &$0$   &$-5$   &$-128$     \\ 
$1D^\prime$ &$-$                &$0$     &$-11$    &$-30$    &$-70$    &$0$    &$0$     &$-5$    &$-2$    &$0$    &$-4$   &$-123$     \\
$1^3D_3$    &$B_{sJ}(6063)$     &$-19$   &$-24$    &$-21$   &$-56$    &$-3$   &$-1$     &$-4$    &$-1$    &$-3$   &$-3$   &$-135$      \\ 
\hline 
\hline
\end{tabular}
\end{table*}

\begin{table*}
\caption{\label{tab:pro} Possibilities (in $\%$) of each coupled channel. } 
\begin{tabular}{cccccccccccccc} 
\hline 
\hline 
State       &                   &$BK$    &$B^*K$  &$BK^*$ &$B^*K^*$  &$B_s\eta$  &$B_s\eta^\prime$ &$B_s^*\eta$ &$B_s^*\eta^\prime$ &$B_s\phi$ &$B_s^*\phi$  &$P_{molecule}$ &$P_{b\bar{s}}$ \\
\hline 
$1^1S_0$    &$B_s$              &$0$     &$2.2$   &$2.0$  &$3.7$     &$0$        &$0$              &$0.4$       &$0.1$      &$0$     &$0.4$    &$8.9$   &$91.1$     \\
$1^3S_1$    &$B_s^{*}$          &$1.0$   &$1.7$   &$1.5$  &$4.8$     &$0.2$      &$0.1$            &$0.3$       &$0.1$      &$0.1$    &$0.3$    &$10.0$ &$90.0$       \\
$1^3P_0$    &$-$                &$23.9$  &$0$     &$0$    &$7.0$     &$1.1$      &$0.1$            &$0$         &$0$       &$0.3$    &$0$     &$32.4$  &$67.6$   \\ 
\hline 
\hline
\end{tabular}
\end{table*}

\begin{table*}
\caption{\label{tab:width1} Decay widths of the $2S$ and $1P$ states.} 
\begin{tabular}{cccccccccc} 
\hline 
\hline 
mode            &$2^1S_0$       &$2^3S_1$              &$1^1P_1$      &$1^3P_1$    &$1^3P_2$  \\
                   &$-$            &$-$                   &$-$           &$B_{s1}(5830)$ &$B_{s2}^*(5840)$ \\
\hline 
$BK$               &$-$            &$34$                  &$-$           &$-$         &$1$[1.01] \\
$B^*K$             &$91$           &$68$                  &$-$           &$0.04$      &$-$[0.08]   \\    
$BK^*$             &$-$            &$-$                   &$-$           &$-$         &$-$   \\ 
$B^*K^*$           &$-$            &$-$                   &$-$           &$-$         &$-$  \\     
$B_s\eta$          &$-$            &$4$                   &$-$           &$-$         &$-$   \\ 
$B_s\eta^\prime$   &$-$            &$-$                   &$-$           &$-$         &$-$  \\
$B_s^*\eta$        &$-$            &$2$                   &$-$           &$-$         &$-$  \\ 
$B_s^*\eta^\prime$ &$-$            &$-$                   &$-$           &$-$         &$-$  \\ 
$B(1^3P_0)K$       &$-$            &$-$                   &$-$           &$-$         &$-$   \\         
$B(1^3P_2)K$       &$-$            &$-$                   &$-$           &$-$         &$-$   \\         
$B(1P)K$           &$-$            &$-$                   &$-$           &$-$         &$-$   \\         
$B(1P^\prime)K$    &$-$            &$-$                   &$-$           &$-$         &$-$   \\         
Total              &$91$           &$108$                 &$-$           &$0.04$      &$1$[1.09]  \\
Exp.               &$-$            &$-$                   &$-$           &$0.5\pm0.4$ &$1.49\pm0.27$ \\
\hline 
\hline
\end{tabular}
\end{table*}

\begin{table*}
\caption{\label{tab:width2}Decay widths of  the $2P$ and $1D$ states.} 
\begin{tabular}{ccccccccc} 
\hline 
\hline
mode           &$2^3P_0$    &$2P$    &$2P^\prime$    &$2^3P_2$    &$1^3D_1$       &$1D$     &$1D^\prime$ &$1^3D_3$ \\ 
                   &$-$         &$-$     &$-$            &$-$         &$B_{sJ}(6158)$ &$-$      &$-$         &$B_{sJ}(6064)$ \\
\hline 
$BK$               &$35$        &$-$     &$-$            &$0.02$      &$28$           &$-$      &$-$       &$12$      \\
$B^*K$             &$-$         &$33$    &$28$           &$1$         &$18$           &$75$     &$34$      &$11$ \\    
$BK^*$             &$-$         &$13$    &$95$          &$53$        &$-$            &$-$      &$-$       &$-$\\ 
$B^*K^*$           &$-$         &$-$     &$-$            &$86$        &$-$            &$-$      &$-$       &$-$ \\     
$B_s\eta$          &$4$         &$-$     &$-$            &$1$         &$10$           &$-$      &$-$       &$0.4$  \\ 
$B_s\eta^\prime$   &$-$         &$-$     &$-$            &$0.1$       &$-$            &$-$      &$-$       &$-$ \\
$B_s^*\eta$        &$-$         &$1$     &$7$            &$3$         &$5$            &$9$      &$1$       &$0.2$ \\ 
$B_s^*\eta^\prime$ &$-$         &$-$     &$-$            &$-$         &$-$            &$-$      &$-$       &$-$  \\  
$B(1^3P_0)K$       &$-$         &$0.1$   &$0.2$          &$-$         &$-$            &$-$      &$-$       &$-$  \\
$B(1^3P_2)K$       &$-$         &$-$     &$-$            &$14$        &$-$            &$-$      &$-$       &$-$  \\
$B(1P)K$           &$-$         &$-$     &$-$            &$1$         &$-$            &$-$      &$-$       &$-$  \\
$B(1P^\prime)K$    &$-$         &$-$     &$-$            &$4$         &$-$            &$-$      &$-$       &$-$  \\
Total              &$39$        &$48$    &$130$          &$165$       &$62$           &$84$     &$35$      &$24$ \\
Exp.               &$-$         &$-$     &$-$            &$-$         &$72\pm18\pm25$            &$-$      &$-$       &$26\pm4\pm4$\\
\hline   
\hline
\end{tabular}
\end{table*}

In this work, we use the nonrelativistic quark model to calculate the mass spectrum of the bottom-strange mesons by considering the coupled channel effects. The model parameters we used are listed in Table~\ref{tab:para}, which have been used to successfully describe the $D$, $D_s$, and $B$ mesons~\cite{Lakhina:2006fy,Li:2010vx,Lu:2016bbk}. The only one unknown parameter $C_{bs}$ is determined to be $C_{bs}=0.169$ by  reproducing the mass of the ground state $B_s$. For the calculation of the decay widths, we used the realistic wave functions by solving the Schrodinger wave equation numerically using the nonrelativistic potential model.
\par
The masses, mass shifts, probabilities, and strong decay widths for the bottom-strange mesons are given in Tables~\ref{tab:spectrum}-\ref{tab:width2}. In the following, we give the explicit discussions.

\subsection{$S$-wave states}
The experimental masses of the lowest-lying bottom-strange mesons $B_s$ ($J^P=0^-$) and $B_s^*$ ($J^P=1^-$) are $5366.93\pm0.10$~MeV and $5415.4\pm1.4$~MeV, which are in good  agreement with our predictions 5367~MeV and 5409~MeV, as shown in Table~\ref{tab:spectrum}. Both of them have no strong decay modes, since their masses are lower than the $BK$ threshold. We could estimate the possibilities of various components, as shown in Table~\ref{tab:pro}. Our results show that the dominant component of the $B_s$ is $b\bar{s}$ ($91.1\%$), while the dominant component of the $B_s^*$ is also $b\bar{s}$ ($90.0\%$). 
\par
There is no experimental information for the $2S$-wave states at present. As shown in Table~\ref{tab:spectrum}, our predicted masses 5949~MeV and 5992~MeV for the $B_s(2^1S_0)$ and $B_s(2^3S_1)$ are reasonable consistent with the results of other models~\cite{Hao:2022ibj,Lu:2016bbk,Godfrey:2016nwn,Ebert:2009ua,Gandhi:2022nnk,Patel:2022hhl}, which indicates that the mass ranges of these two states should be $5947\sim 6025$~MeV and $5972\sim 6033$~MeV, respectively. Furthermore, we have shown the strong decay behaviors of these two states in Table~\ref{tab:width1}. The $B_s(2^1S_0)$ can only decay into $B^*K$ with the predicted width of 91~MeV, and the $B_s(2^3S_1)$ can decay into $BK$ (34~MeV) and $B^*K$ (68~MeV) with total width of $108$~MeV. The ratio between two decay modes is
\begin{equation}
\frac{\Gamma(B_s(2^3S_1)\to B^{*+}K^-)}{\Gamma(B_s(2^3S_1)\to B^+K^-)} =2.0,
\end{equation} 
which is in good agreement with the results of Refs.~\cite{Sun:2014wea,Lu:2016bbk}.

\subsection{$P$-wave states}
The $B_{s1}(5830)$ and $B_{s2}^*(5840)$ are identified as the $1^+$ and $2^+$ states. According to Table~\ref{tab:spectrum},  the predicted masses of the $B_s(1^3P_0)$ and $B_s(1P)$ are below the $BK$ threshold, therefore these states has no strong decay modes~\cite{Lu:2016bbk,Zeng:1994vj,Sun:2014wea}. As shown in Table~\ref{tab:pro}, our results suggest that the dominant components of the $B_s(1^3P_0)$ are  $BK$ ($23.9 \%$) and  $b\bar{s}$ ($67.6 \%$).
With the mixing angle $-55.8^\circ$, the predicted masses of the $B_s(1P^\prime)$ and $B_s(1^3P_2)$ are 5805~MeV and 5825~MeV, which are reasonable consistent with those of the $B_{s1}(5830)$ and $B_{s2}^*(5840)$, respectively. Meanwhile, both the decay widths for these two states are very narrow, consistent with the experimental data. On the other hand, the decay ratio between two decay modes is
\begin{equation}
\frac{\Gamma(B_{s2}^*(5840)\to B^{*+}K^-)}{\Gamma(B_{s2}^*(5840)\to B^+K^-)} = 0.096,
\end{equation} 
which is also consistent with the LHCb data of  $0.093\pm0.013\pm0.012$~\cite{LHCb:2012iuq}. Thus, one could conclude that the $B_{s1}(5830)$ and $B_{s2}^*(5840)$ should be the $B_s(1P^\prime)$ and $B_s(1^3P_2)$, respectively, same as discussed in Ref.~\cite{Lu:2016bbk}.
\par
As shown in Table~\ref{tab:spectrum} and Table~\ref{tab:width2}, the masses of the $2P$-wave states are predicted to be 6214~MeV, 6274~MeV, 6307~MeV, and 6352~MeV, respectively, and their total decay widths are 39~MeV, 48~MeV, 130~MeV, and 165~MeV, respectively. Our results show that the $B_s(2^3P_0)$ mainly decays into $BK$, while the main decay modes of the $B_s(2^3P_2)$ state are $BK^*$ and $B^*K^*$. With mixing angle $-54.5^\circ$, both the $B_s(2P)$ and $B_s(2P^\prime)$ mainly decay to $B^*K$ and $BK^*$. 

\subsection{$D$-wave states}
According to Table~\ref{tab:spectrum} and Table~\ref{tab:width2}, our predicted mass and width of the $B_s(1^3D_1)$ is 6157~MeV and 62~MeV, consistent with the mass $6158\pm4\pm5$~MeV and width $72\pm18\pm25$~MeV of the $B_{sJ}(6158)$, which implies that the state $B_{sJ}(6158)$ could be regarded as the candidate of the $B_s(1^3D_1)$. Our predicted value for the mass and decay width for $B_s(1^3D_3)$ are 6079~MeV and 24~MeV, consistent with  the mass $6063.5\pm1.2\pm0.8$~MeV and width $26\pm4\pm4$~MeV of the $B_{sJ}(6064)$, respectively, which implies that the $B_{sJ}(6064)$ could be considered as a candidate of the $B_s(1^3D_3)$.

As for the two $J^P=2^+$ states, they are mixing states due to the spin-orbit mixing. According to Table~\ref{tab:spectrum}, the predicted masses of the  $B_s(1D)$ and $B_s(1D^\prime)$ are $6077$~MeV and $6154$~MeV, respectively, which are in reasonable agreement with the predictions of Refs.~\cite{Lu:2016bbk,Hao:2022ibj}. Their decay properties are given in Table~\ref{tab:width2}, which show that the $B_s(1D)$ and $B_s(1D^\prime)$ can decay to $B^*K$ and $B_s^*\eta$, and the dominant decay channel for both states is $B^*K$. 

On the other hand, based on a nonrelativistic linear potential model, the $B_{sJ}(6064)$ could be explained as $B_s(1^3D_3)$ with predicted mass $M=6067$~MeV and width $\Gamma=13$~MeV. In Ref.~\cite{li:2021hss}, the $B_{sJ}(6064)$ was considered as $B_s(2^3S_1)$, using the heavy quark model effective theory, but the predicted width $\Gamma=170\pm1.5$~MeV is too larger than the experimental data of $26\pm4\pm4$~MeV.

\section{Summary}
\label{sec:summary}
In this work, we have systematically investigated the mass spectrum and the strong decay properties of the bottom-strange mesons in the coupled channel model with a nonrelativistic potential. The wave functions are obtained by solving the Schrodinger wave equation using the Gaussian expansion method, and used to calculate the strong decay widths with the $^3P_0$ model.

We could assign the $B_{s1}(5830)$ and $B_{s2}^*(5840)$ as $B_s(1P^\prime)$ and $B_s(1^3P_2)$, respectively. 
Furthermore, our results suggest that the $B_{sJ}(6158)$ and $B_{sJ}(6064)$ could be regarded as candidates for the $B_s(1^3D_1)$ and $B_s(1^3D_3)$ states, respectively, both the $B_{sJ}(6064)$ and $B_{sJ}(6158)$ mainly decay to the $BK$ and $B^*K$ channels. 
\par
Specifically, our results show that the dominant component of the ground states $B_s(1^1S_0)$ and $B_s(1^3S_1)$ is $b\bar{s}$ (91.1\% and 90.0\%),  and the $b\bar{s}$ component of the $B_s(1^3P_0)$ is about 67.6\%.

Furthermore, we also calculated the masses and decay widths of the $2S$ states. The main decay mode of $B_s(2^1S_0)$ is $BK$ with a total width of 91 MeV, and $B_s(2^3S_1)$ can decay to $BK$ (34 MeV) and $B^*K$ (68 MeV), which will be helpful in searching for them in the experiments.

\section{Acknowledgements}
E.Wang acknowledge the support from the National Key R\&D Program of China (No. 2024YFE0105200).
This work is supported by the Natural Science Foundation of Henan under Grant No. 232300421140 and No. 222300420554, the National Natural Science Foundation of China under Grant No. 12475086 and No. 12192263.

\bibliographystyle{unsrt}
\bibliography{cite}  

\end{document}